\documentclass[12pt]{article}
\def\beq{\begin{equation}}
\def\eeq{\end{equation}}
\def\bea{\begin{eqnarray}}
\def\eea{\end{eqnarray}}
\def\bq{\begin{quote}}
\def\eq{\end{quote}}
\def\csumA{\sum_{A=1}^{N}}
\def\csumB{\sum_{B=1}^{N}}
\def\csumAB{\sum_{A, B=1}^{N}}

\def\gappeq{\mathrel{\rlap {\raise.5ex\hbox{$>$}}
{\lower.5ex\hbox{$\sim$}}}}

\def\lappeq{\mathrel{\rlap{\raise.5ex\hbox{$<$}}
{\lower.5ex\hbox{$\sim$}}}}

\begin{document}
\topmargin -0.5cm
\oddsidemargin -0.3cm
\pagestyle{empty}
\begin{flushright}
{HIP - 2010 - 04/TH}
\end{flushright}
\vspace*{5mm}
\begin{center}
\large
{\bf On the compatibility of non-holonomic systems and certain related variational systems}\footnote{Presented at the 4th International Young Researchers Workshop on Geometry, Mechanics and Control, January 11-13, 2010 - Ghent, Belgium.}   
\\
\normalsize
\vspace*{1.5cm} 
{\bf Christofer Cronstr\"{o}m}$^{*)}$ \\
\vspace{0.3cm}
Helsinki Institute of Physics\\
P. O. Box 64, FIN-00014 University of Helsinki, Finland \\
 
\vspace*{3cm}

{\bf ABSTRACT} \\

\end{center}

I consider the equations of motion which follow from d'Alembert's principle for a general mechanical system in a space of $N$ dimensions,  constrained by a non-holonomic constraint which is linear and homogeneous in the generalised velocities. The variational equations of motion which follow for the same system by assuming the validity of  a specific variational action principle, in which  the non-holonomic constraint is implemented by means of the multiplication rule in the calculus of variations are also considered.  It is shown that these two types of equations of motion are not compatible in a space of dimension $N \geq 3$ if the constraint is genuinely non-holonomic. This means that these two types of equations of motion do not have coinciding general solutions.

\vspace*{5mm}
\noindent

\vspace*{1cm} 
\noindent

\noindent 
$^{*)}$ e-mail address: Christofer.Cronstrom@Helsinki.fi \\

\noindent
45.50.Pk, 45.10.Na, 02.30.Xx
\newpage

\pagestyle{plain}
\pagenumbering{arabic}
\setcounter{page}{1}

\section{Introduction}

Hamilton's principle for mechanical systems with non-holonomic constraints has  recently been discussed by Flannery \cite{Flannery}.  In particular a variational formulation of the equations of motion of a mechanical system was discussed both for holonomic and non-holonomic constraints. It was shown  that while the equations of motion for a system with holonomic constraints can be obtained as variational equations,  with the  constraints being taken into account by the multiplication rule in the calculus of variations \cite{Mikhlin},  the corresponding procedure with non-holonomic constraints leads to equations which differ  from the equations of motion which follow from the well-known principle of d'Alembert.  The questions discussed by Flannery are by no means new; they have been discussed
in the literature in several papers. For a selection of references other than those given by Flannery, I refer to some of the references contained in a recent paper\footnote{There are some regrettable misprints in this paper. The equations (17), (26) and (34) in Ref. \cite{CCTRJMP} contain the redundant symbols = 0 at the end of the equations.} by Cronstr\"{o}m and Raita \cite{CCTRJMP}.

Even though the equations of motion following from the principle of d'Alembert and from the variational action principle with non-holonomic constraints are different in form, there is still the possibility that the equations in question may have the same solutions. It was demonstrated by Pars \cite{Pars}, that  this is not the case for a particular example with a specific non-holonomic constraint in a three-dimensional space. As such, this validates the assertion that the solutions to the two different types of equations of motion are different in general, at least in a space of three dimensions.  It has only recently been proved,  \cite{CCTRJMP}, \cite{CCMN70}, that this is also the case generally in configuration spaces of dimension $N \geq 3$. Below I discuss an improved  version of the  proofs in \cite{CCTRJMP} and \cite{CCMN70}, separately  for $N= 3$ and $N\geq4$.

I consider a general autonomous system with a finite number of degrees of freedom, restricted only by reasonable smoothness conditions.  For simplicity I consider  only the case of one non-holonomic  constraint, which is taken to be linear and homogeneous in the  generalised velocities of the system. 
In principle the proof is valid when the configuration space of the system is a fairly general smooth $N$-dimensional manifold, with $N \geq 3$. However, since the argumentation in the proofs of this paper  for the most part is local, it is sufficient to use a co-ordinate formulation and consider only one appropriately chosen co-ordinate patch $D_{q}$, with local co-ordinates designated as $q = (q^{1},...,q^{N})$.

In the case of two-dimensional systems there is nothing to prove, since every non-holonomic constraint of the kind considered in this paper  can be reduced to an equivalent holonomic constraint when $N=2$.

The proof given in this paper implies definitely that the standard variational principle, which is valid for unconstrained systems, and which can be  generalised to cover the case of holonomic constraints by using the multiplication rule in the calculus of variations, is not in general consistent with the principle of d'Alembert, if  one uses the multiplication rule with a genuinely non-holonomic constraint. 

\section{The variational principles}

Consider an autonomous  mechanical system with  generalised co-ordinates 
$q = (q^{1},...,q^{N})$, and  velocities $\dot{q} = (\dot{q}^{1},...,\dot{q}^{N})$. Throughout this paper I assume the the variables $q = (q^{1},...,q^{N})$ belong to a chosen co-ordinate patch $D_{q}$, which is a contractible domain in configuration space. 

The kinetic energy of the system  is denoted by by  $T$, and the generalised applied forces acting on the system are denoted by  $Q_{A}, A = 1,...,N$.  It will further be assumed that the generalised forces can be
expressed in terms of a generalised potential $V(q, \dot{q})$, 
\beq
\label{GenV}
Q_{A} = - \frac{\partial V}{\partial q^{A}} +  \frac{d}{dt}\left (\frac{\partial V}{\partial  \dot{q}^{A}}\right ), A = 1,...,N.
\eeq
It is appropriate to introduce a Lagrange function $L$ as follows,
\beq
\label{lagrangian}
L(q, \dot{q}) := T (q, \dot{q}) - V(q, \dot{q}).
\eeq
If the variables $(q^{1},...,q^{N})$ are not constrained in any way, then, as is well known,  the equations of motion of the system follow from the following variational principle,
\beq
\label{actionprS}
\delta S = 0,
\eeq
where
\beq
\label{defS}
S := \int dt\,L(q, \dot{q}).
\eeq
The equations of motion are the following Euler-Lagrange equations,
\beq
\label{ELeq}
\frac{d}{dt}\left (\frac{\partial L(q, \dot{q})}{\partial  \dot{q}^{A}}\right ) - \frac{\partial L(q, \dot{q})}{\partial q^{A}}   =   0 ,\; A  =  1, \ldots, N.
\eeq

I then consider a situation in which the variables  $(q^{1},...,q^{N})$ are constrained by a condition of the form
\beq
\label{holonomic}
\Psi(q^{1},...,q^{N}) = C,
\eeq
where $\Psi(q)$ is a smooth (continuously differentiable) function of its variables and $C$ is a constant. The condition (\ref{holonomic}) is a {\em holonomic} constraint. It is well-known that the equations of motion for the system also in this case can be obtained from the variational principle $\delta S = 0$,  under the constraint (\ref{holonomic}). Using the multiplication rule in the calculus of variations \cite{Mikhlin} this leads to the following condition,
\beq  
\label{mprule1}
\delta \int \,dt \left [L(q, \dot{q}) +  \lambda\, \Psi(q) \right ] = 0,
\eeq
which involves a Lagrange multiplier $\lambda$.
The Euler-Lagrange equations following from Eqn.  (\ref{mprule1}) are as follows
\beq
\label{mpruleEq}
\frac{d}{dt}\left (\frac{\partial L(q, \dot{q})}{\partial  \dot{q}^{A}}\right ) - \frac{\partial L(q, \dot{q})}{\partial q^{A}}   =  \lambda\, \frac{\partial \Psi(q)}{\partial q^{A}},\; A  =  1, \ldots, N,
\eeq
The equations (\ref{mpruleEq}) constitute a set of $N$ second-order differential equations, which, under appropriate boundary conditions, and together with the constraint (\ref{holonomic}), are supposed to determine the quantities $q^{1}, q^{2}, \ldots, q^{N}$, as well as the Lagrange multiplier $\lambda$.

I then finally consider the case of a {\em non-holonomic} constraint, which is taken to be a linear and homogeneous in the generalised velocity variables $\dot{q}^{1},...,\dot{q}^{N}$. The non-holonomic constraint is thus of the following form,
\beq
\label{1nonhol}
\sum_{A=1}^{N} a_{A}(q)\,\dot{q}^{A} = 0,
\eeq
where the $N$ quantities $a_{A}(q), A = 1, 2, \ldots, N$ are given smooth (continuously differentiable) functions, which transform as covariant vector components under co-ordinate transformations. 

A metric $g$ is involved in the notation in Eq. (\ref{1nonhol}), which could also be written as follows,
\beq
\label{gAB}
\sum_{A, B=1}^{N}g_{AB}(q)\,a^{A}(q)\,\dot{q}^{B} = 0.
\eeq
Tensor-indices are thus lowered or raised with the metric tensor $g_{AB}(q)$ and $g^{AB}(q)$, respectively, where
\beq
\label{riseg}
\csumB g_{AB}(q)\,g^{BC}(q) = g_{A}^{~~C} :=    \left \{ \begin{array}{ll}
1 & \mbox{if $A = C$}\\
0 & \mbox{if $A \neq C$,}
\end{array}
\right .
\eeq
for $A, C = 1, 2, \ldots, N$.

It is clear that that not every component of $a_{A}(q)$ can vanish identically in $D_{q}$, since otherwise there would not be any non-holonomic condition at all in the problem.

The question is then whether one can use the analogue of the equations (\ref{mprule1}) for a non-holonomic constraint of the form (\ref{1nonhol}), simply by enforcing the constraint (\ref{1nonhol}) by means of the  multiplication rule throughout a domain which is supposed to contain all the competing paths in the variational principle. 

I now prefer to use the special notation
\beq
\label{ell0}
L_{0}(q, \dot{q}) = T(q, \dot{q}) - V(q, \dot{q}). 
\eeq
This special notation is used as a reminder of the fact that the quantity $L_{0}(q,\dot{q})$ that one should be aware of the potential danger involved in using the quantity $L_{0}$  as a Lagrangian in Hamilton's principle when the constraints are genuinely  non-holonomic.  In the absence of a  constraint of the form (\ref{1nonhol}),  the quantity $L_{0}(q, \dot{q})$ would naturally be the Lagrangian of the system.  

The following condition is a straightforward analogue of the variational  condition (\ref{mprule1}),
\beq  
\label{varnh3}
\delta \int \,dt \left [L_{0}(q, \dot{q}) -  \mu\csumA \,a_{A}(q)  \dot{q}^{A} \right ] = 0,
\eeq
which involves a Lagrange multiplier $\mu$. The variational principle underlying the condition (\ref{varnh3}) is, as such, a possible variational principle for certain types of problems, but it does not necessarily give rise to appropriate equations of motion in Newtonian classical mechanics. 

The condition  (\ref{varnh3}) gives rise to the following Euler-Lagrange equations,
\beq
\label{CC1}
\frac{d}{dt}\left (\frac{\partial L_{0}(q, \dot{q})}{\partial  \dot{q}^{A}}\right ) - \frac{\partial L_{0}(q, \dot{q})}{\partial q^{A}}   =   \dot{ \mu} \,a_{A}(q)  + \mu\, \sum_{B =1}^{N} \;M_{AB}(q) \dot{q}^{B},\; A  =  1, \ldots, N,
\eeq
where the quantities $M_{AB}(q), A, B = 1. 2. \ldots, N$ are given as follows,
\beq
\label{matrixM}
M_{AB}(q)  :=  \frac{\partial a_{A}(q)}{\partial q^{B}} - \frac{\partial a_{B}(q)}{\partial q^{A}}, \; A, B = 1, \ldots, N.
\eeq
The variational equations of motion (\ref{CC1}) are precisely those which have been proposed from time to time in the literature, as appropriate variational equations of motion  for mechanical systems with non-holonomic constraints of the type considered in this paper.

It should be noted that the equations (\ref{CC1}) and the constraint (\ref{1nonhol}), admit a first integral for the system,
\beq
\label{energyint}
E := \sum_{A=1}^{N}\, \dot{q}^{A}\, \frac{\partial L_{0}(q, \dot{q})}{\partial  \dot{q}^{A}} - L_{0}(q, \dot{q}),
\eeq
where $E$ is the constant energy of the system.

Consider now the special case in which the constraint (\ref{1nonhol}) is integrable. This means that there exists a function $\Psi(q)$, say, such that
\beq
\label{integr}
a_{A}(q) = \frac{\partial}{\partial q^{A}} \Psi(q), A = 1, 2, \ldots, N.
\eeq
Then the constraint (\ref{1nonhol}) is in fact a  holonomic constraint of the kind given in Eq.
 (\ref{holonomic}).

Inserting the expressions (\ref{integr}) in the equations (\ref{CC1}) one obtains the following equations,
\beq
\label{CC1hol}
\frac{d}{dt}\left (\frac{\partial L(q, \dot{q})}{\partial  \dot{q}^{A}}\right ) - \frac{\partial L(q, \dot{q})}{\partial q^{A}}   =   \dot{ \mu} \, \frac{\partial \Psi(q)}{\partial q^{A}},\; A  =  1, \ldots, N,
\eeq
which naturally agree with the equations (\ref{mpruleEq}) when one makes the change of notation $\dot{\mu} \rightarrow \lambda$.

\section{Genuinely non-holonomic constraints}
\label{seconstr}

It was noted above that if there exists a function $\Psi(q)$ such that the condition (\ref{integr}) holds true, then the constraint (\ref{1nonhol}) is equivalent to a holonomic constraint of the  kind given in Eq.
 (\ref{holonomic}). If the condition (\ref{integr}) holds true, then necessarily,
 \beq
\label{closeM}
M_{AB}(q) = 0, \;\;A, B = 1, 2, \ldots, N, \; q \in D_{q},
\eeq
where the quantities $M_{AB}, A, B = 1, 2, \ldots, N$ were defined in Eqns. (\ref{matrixM}). Conversely, in a contractible domain, the integrability conditions (\ref{closeM}) are also sufficient for the integrability of the constraint (\ref{1nonhol}). Thus, the constraint (\ref{1nonhol}) is integrable, and therefore equivalent to a holonomic constraint, if and only if the integrability conditions (\ref{closeM}) are fulfilled.

In the case at hand, the constraint (\ref{1nonhol}) is equivalent to the following constraint,
\beq
\label{intfact}
\Phi(q)\, \sum_{A=1}^{N}\,a_{A}(q)\,\dot{q}^{A} =  0,
\eeq
where $\Phi(q)$ is some appropriate smooth function, which does not vanish identically. It may be that one can choose the function $\Phi(q)$ in (\ref{intfact}) in such a manner that this constraint is integrable.
If that is the case one says that $\Phi(q)$ is an integrating factor. Replacing $a_{A}$ by $\Phi\,a_{A}, A = 1, 2, \ldots, N$ in the integrability conditions (\ref{closeM}), one finds the following necessary conditions for the existence of an integrating factor,
\beq
\label{intcomp}
a_{A}(q)M_{BC}(q) + a_{B}(q)M_{CA}(q) + a_{C}(q)M_{AB}(q) = 0, \, A, B, C = 1,2,\ldots, N.
\eeq
The number of independent  integrability conditions  of the type (\ref{intcomp}) is  $N_{c}$, where \cite{Ince},  
\beq
\label{numbint}
N_{c} = \frac{1}{2}(N-1)(N-2).
\eeq
In three-dimensional space ($N = 3$) there is thus only one integrability condition of the type 
(\ref{intcomp}), namely the following condition,
\beq
\label{3dim}
a_{1}(q)M_{23}(q) + a_{2}(q)M_{31}(q) + a_{3}(q)M_{12}(q) = 0.
\eeq
In the case of a contractible domain the integrability conditions (\ref{intcomp}) are also sufficient for the existence of an integrating factor $\Phi(q)$.

Thus, if there exists an integrating factor $\Phi(q)$, then the original constraint (\ref{1nonhol}), which has the appearance of a non-holonomic constraint, can be replaced by an equivalent holonomic constraint 
by making use of the integrating factor.

In what follows it is important to consider only  such constraints which are {\em genuinely non-holonomic}. A genuinely non-holonomic constraint, linear and homogeneous in the velocity components $\dot{q}^{A}, A = 1, 2, \ldots, A$, is a constraint of the form (\ref{1nonhol})  which is neither integrable as such, nor integrable by means of an integrating factor. Thus, for a genuinely non-holonomic constraint neither the integrability conditions (\ref{closeM}) nor the integrability conditions (\ref{intcomp})
are fulfilled.

\section{The d'Alembert equations with  constraints}

The principle of d'Alembert (see  e.g.  the classical texts by Goldstein \cite{Goldstein59}  or Whittaker \cite{Whittaker}) 
 gives the following equation,
\beq
\label{Lag0}
\sum_{A=1}^{N} \left \{\frac{d}{dt}\left (\frac{\partial T}{\partial  \dot{q}^{A}} \right ) - \frac{\partial T}{\partial q^{A}} - Q_{A}\right \} \delta q^{A} = 0,
\eeq
where the quantities $\delta q^{A}$ are  {\em virtual displacements} of the system. The meaning of the symbols $T$ and $Q_{A}$, respectively, is the same as before, {\em i.e.} $T$ is the kinetic energy and $Q_{A}$ is the $A$:th  component of the generalised external force imposed on the system. If the virtual displacements $\delta q^{A}, A = 1,\ldots, N$, are independent, then  Eq.  (\ref{Lag0}) results in the ordinary d'Alembertian  equations of motion,
\beq
\label{Lag1}
\frac{d}{dt}\left (\frac{\partial T}{\partial  \dot{q}^{A}} \right ) - \frac{\partial T}{\partial q^{A}} = Q_{A}, \; A = 1, \ldots, N.
\eeq

Consider the generalisation of  the discussion above to systems with a  constraint of the  form (\ref{1nonhol}) given above. The derivation given below of the equations of motion for this system, with a potentially genuinely non-holonomic constraint, can be found  in the textbook by  Whittaker \cite{Whittaker}.

Implement the constraint (\ref{1nonhol}) by regarding the system to be acted on by both the external applied forces  $Q_{A}, \, A=1,\ldots,N$, and  by certain additional forces of constraint $Q'_{A},\, A=1,\ldots,N$, which force the system to satisfy the condition (\ref{1nonhol}). Equation~(\ref{Lag0}) is then replaced by the following equation,
\beq
\label{Lag3}
\sum_{A=1}^{N}\,\left \{\frac{d}{dt}\left (\frac{\partial T}{\partial  \dot{q}^{A}} \right ) - \frac{\partial T}{\partial q^{A}} - Q_{A} - Q'_{A} \right \} \delta q^{A} = 0,
\eeq
In Eq. (\ref{Lag3})  the virtual displacements $\delta q^{A}, A = 1, \ldots, N$,  can now be regarded as independent.  Thus one obtains  the equations of motion,
\beq
\label{Lagnh}
\frac{d}{dt}\left (\frac{\partial T}{\partial  \dot{q}^{A}} \right ) - \frac{\partial T}{\partial q^{A}} =Q_{A} + Q'_{A}, \,A = 1, \ldots, N.
\eeq
The forces of constraint, $Q'_{A}, A = 1, \ldots, N$, are {\em a priori} unknown, but they are such that, in any instantaneous displacement $\delta q^{A}, A = 1,\ldots, N$, consistent with the constraint (\ref{1nonhol}), they do no work.  The non-holonomic constraint (\ref{1nonhol})  implies the following condition on the possible instantaneous displacements   $\delta q^{A}, A = 1,\ldots, N$,
\beq
\label{cdelta}
\sum_{A=1}^{N}\,a_{A}(q) \delta q^{A} = 0.
\eeq
For any instantaneous displacements $\delta q^{A}, A = 1,\ldots, N$, which satisfy the condition (\ref{cdelta}), the work $\delta W'$ done by the constraint forces $Q'_{A}, A=1,\ldots,N$ must be equal to zero,  {\em i.e.},
\beq
\label{workc}
\delta W'  := \sum_{A=1}^{N} Q'_{A}\delta q^{A} = 0.
\eeq
The conditions (\ref{cdelta}) and (\ref{workc})  together imply that
\beq
\label{cforce}
Q'_{A} =  \lambda a_{A}(q),\, A = 1,\ldots, N,
\eeq
where the quantity $\lambda$ is a time-dependent parameter to be determined.  Eqns. (\ref{Lagnh}) have  thus been reduced to the following form,
\beq
\label{Lagnh2}
\frac{d}{dt}\left (\frac{\partial T}{\partial  \dot{q}^{A}} \right ) - \frac{\partial T}{\partial q^{A}} =Q_{A} + 
 \lambda a_{A}(q), \,A = 1, \ldots, N.
\eeq
These $N$  equations of motion are  consequences of the principle of  d'Alembert. They will be called {\em d'Alembertian equations}. One should still add the  equation of constraint (\ref{1nonhol}) to the equations of motion above. There are thus altogether  $N+ 1$ equations for the determination of $N+1$ quantities $q^{A}(t), A = 1,\ldots,N$, and $\lambda(t)$, when appropriate boundary conditions for the quantities $q^{1},...,q^{N}$ and $\dot{q}^{1},...,\dot{q}^{N}$ are given.

One should observe that in the argument above, it is \underline{not}  required that the constraint equation (\ref{1nonhol})  be in force under general  variations $q_{j} \rightarrow q_{j} + \delta q_{j}$; the constraint (\ref{1nonhol}) is \underline{only} imposed on the actual motion of the system. 

Assuming that the generalised force components $Q_{A}, A = 1, 2, \ldots, N$ can be expressed in terms of a generalised potential $V(q, \dot{q})$, as in Eq. (\ref{GenV}), one can rewrite the d'Alembertian equations  (\ref{Lagnh2})  as follows,
\beq
\label{Lagnh3}
\frac{d}{dt}\left (\frac{\partial L_{0}(q, \dot{q})}{\partial  \dot{q}^{A}}\right ) - \frac{\partial L_{0}(q, \dot{q})}{\partial q^{A}} =  \lambda  a_{A}(q), \;\,A = 1, \ldots, N,
\eeq
where the quantity $L_{0}(q, \dot{q})$ has been defined in Eq. (\ref{ell0}).

The equations (\ref{Lagnh3}) will also be referred to as d'Alembertian equations in what follows.

The d'Alembertian equations of motion (\ref{Lagnh3}) admit a first integral, namely the expression (\ref{energyint}). This is the same first integral as the one  obtained from the variational equations (\ref{CC1}). 

\section{Non-equivalence of the principle of d'Alembert and the variational action principle with genuinely non-holonomic constraints}

I now consider the d'Alembertian equations of motion (\ref{Lagnh3}) and the variational equations (\ref{CC1}), respectively,  as {\em initial value} problems. This means that the co-ordinate patch in which the 
quantities $q^{1}, q^{2}, \ldots, q^{N}$ are local co-ordinates, is assumed to be chosen so that it includes the initial value point $q_{0} = (q_{0}^{1}, q_{0}^{2}, \ldots, q_{0}^{N})$. 

The question is whether the equations (\ref{Lagnh3}) and (\ref{CC1}), respectively, can have the same solutions for $q^{1}(t), \ldots, q^{N}(t)$ in general, despite the fact that these equations are not identical. It will be shown below that this is not the case. 

Naturally, the functions $L_{0}(q, \dot{q})$ and $a_{A}(q)$ entering into the equations (\ref{Lagnh3})
and (\ref{CC1}), respectively, will have to satisfy appropriate smoothness conditions in order that these equations may have solutions. I will not enter into a discussion of such smoothness conditions, but rather assume that the equations (\ref{Lagnh3}) and (\ref{CC1}), respectively, have {\em e.g.} $C^{2}$-solutions in some appropriate time-interval, for given initial values for the co-ordinates $q^{A}$ in the domain $D_{q}$ and for given initial values for the velocities $\dot{q}^{A}, A = 1, \ldots, N$,
\beq
\label{initq}
\left [q^{A}(t)\right ]_{t=t_{0}} = q_{0}^{A},\; \left [\dot{q}^{A}(t)\right ]_{t=t_{0}} = \dot{q}_{0}^{A},\; A = 1, \ldots, N.
\eeq
The initial values $ q_{0}^{A}$ and  $\dot{q}_{0}^{A}$ at $t = t_{0}$ are free parameters  within an appropriate region of the configuration- and velocity space, except for the restriction
\beq
\label{initrestr}
\sum_{A=1}^{N} \,a_{A}(q_{0})\, \dot{q}_{0}^{A} = 0.
\eeq
The condition (\ref{initrestr})  is a consequence  of  the non-holonomic constraint (\ref{1nonhol}). 

After these considerations concerning the initial values I will prove a theorem on the incompatibility of the  d'Alembertian and variational equations of motion, respectively.

\subsection{The incompatibility theorem}
\label{incomp}

Consider the d'Alembertian and variational equations of motion (\ref{Lagnh3}) and (\ref{CC1}), respectively, which involve the genuinely non-holonomic constraint (\ref{1nonhol}). If $N=3$ the d'Alembertian and variational equations of motion are not compatible in the sense that they do not have coinciding solutions. If $N\geq4$ the  d'Alembertian and variational equations of motion are not compatible in the sense that they do not have coinciding solutions with arbitrary general initial values  (\ref{initq}), which satisfy the condition (\ref{initrestr}).
    
It is  assumed that the d'Alembertian equations (\ref{Lagnh3})  have unique smooth  ({\em e.g.} $C^{2}$) solutions $(q^{1}(t), \ldots, q^{N}(t))$ in an appropriate time-interval, satisfying a genuinely non-holonomic constraint of the type (\ref{1nonhol}), and the general initial value conditions (\ref{initq}) and (\ref{initrestr}).  I call such solutions general solutions in what follows. 

The method of proof is by {\em reductio ad absurdum}, {\em i.e.}, one makes the assumption that   the variational equations (\ref{CC1})  have solutions which coincide with the  general solutions of the d'Alembertian equations (\ref{Lagnh3}), and shows that this assumption leads to contradictions. 

The proof given below is designed in such a manner that it is  independent of any specific properties of the kinetic energy $T$ and generalised potential $V$, respectively,  for the problem under consideration.

\subsection{Preliminary considerations}
\label{preliminary}

Assume now that the equations (\ref{CC1}) and (\ref{Lagnh3}) have coincident solutions. By subtracting Eqns. (\ref{CC1}) from Eqns. (\ref{Lagnh3}), one obtains the following equations,
\beq
\label{diffdACC1}
(\lambda - \dot{\mu})\,a_{A}(q) = \mu \; \sum_{B=1}^{N} M_{AB}(q)\,\dot{q}^{B} = 0, \; A = 1, \ldots, N,
\eeq
which  have to be satisfied by the general solutions $(q^{1}, \ldots, q^{N})$ of the d'Alembertian
equations (\ref{Lagnh3}). 

It should be noted that one must necessarily have $\mu \not\equiv 0$ in the equations  (\ref{diffdACC1}) above, since otherwise the variational equations (\ref{CC1}) would contain no reference whatsoever to the constraints (\ref{1nonhol}).  Further, if $\mu \equiv 0$, it follows from the equations (\ref{diffdACC1})
that also the following conditions must hold true,
\beq
\label{exceptA}
\lambda\,a_{A} \equiv 0,\; A =  1. \ldots, N.
\eeq
The conditions (\ref{exceptA}) above imply that  either 
\beq
\label{lamndanoll}
\lambda \equiv 0,
\eeq
or
\beq
\label{anoll}
a_{A}(q) \equiv 0,\; A =  1. \ldots, N.
\eeq
If the conditions (\ref{anoll}) are true, there are no constraints to be considered so the the basic question under consideration disappears. Finally, if  the condition (\ref{lamndanoll}) holds true in addition to the condition $\mu \equiv 0$, then the d'Alembertian equations of motion (\ref{Lagnh3}) and the  variational equations of motion (\ref{CC1}) are trivially identical, so the the basic question under consideration disappears also in that case. Hence one must have $\mu \not\equiv 0$.

Using the  notation
\beq
\label{Gamma}
\Gamma := \frac{\lambda - \dot{\mu}}{\mu}.
\eeq
\noindent
one can write the conditions (\ref{diffdACC1}) in an  equivalent form as follows,
\beq
\label{eqnsM}
\sum_{B=1}^{N} M_{AB}(q)\,\dot{q}^{B} = \Gamma\,a_{A}(q), \; A = 1, \ldots, N,
\eeq
The tensor equations (\ref{eqnsM}) constitute a set of $N$ linear algebraic equations in the variables 
$\dot{q}^{A}, A = 1, \ldots, N$, for any fixed value of $q := (q^{1}, q^{2}, \ldots, q^{N}) \in D_{q}$.  It should be noted that the quantities $M_{AB}(q)$ in Eqns. (\ref{eqnsM}) are anti-symmetric upon an interchange $A \leftrightarrow B$ of the indices $A$ and $B$,
\beq
\label{antisym}
M_{AB}(q) = - M_{BA}(q), \; A, B = 1, 2, \ldots, N.
\eeq
Thus the matrix-like quantity $M(q) := (M_{AB}(q))$ is {\em skew}.

\subsection{The case $N = 3$}
\label{dimension3}

I consider the case N = 3 separately, since in that case the analysis of the tensor equations 
(\ref{eqnsM}) is particularly simple. Thus, for $N = 3$ the equations (\ref{eqnsM}) read as follows,
\beq
\label{eqnsM3}
\sum_{B=1}^{3} M_{AB}(q)\,\dot{q}^{B} = \Gamma\,a_{A}(q), \; A = 1, 2, 3.
\eeq
It should be noted that since $N = 3$ and $M(q)$ is skew, one necessarily has
\beq
\label{detM3}
\det M(q) = 0.
\eeq

The quantity $\Gamma$ in Eqns. (\ref{eqnsM3}) is a so far unknown parameter, but one must have either $\Gamma \equiv 0$ or $\Gamma \not\equiv 0$,

Consider first the case $\Gamma \equiv 0$.  Then the velocity components $(\dot{q}^{1}, \dot{q}^{2}, \dot{q}^{3})$
satisfy the homogeneous equations
\beq
\label{homM3}
\sum_{B=1}^{3} M_{AB}(q)\,h^{B}(q) = 0, \; A = 1, 2, 3.
\eeq
The homogeneous equations (\ref{homM3}) have the following non-trivial solutions,
\beq
\label{solM3}
(h^{1}(q), h^{2}(q), h^{3}(q)) := \alpha ( M_{23}(q), M_{31}(q), M_{12}(q)), \; \alpha \not\equiv  0.
\eeq
However, the solutions (\ref{solM3}) for $(\dot{q}^{1}, \dot{q}^{2}, \dot{q}^{3})$ must also satisfy the constraint (\ref{1nonhol}) with $N=3$, {\em i.e.},
\beq
\label{const3}
\sum_{A=1}^{N} a_{A}(q)\, \dot{q}^{A} = 0.
\eeq
This constraint then implies that
\beq
\label{acond}
a_{1}(q)M_{23}(q) + a_{2}(q)M_{31}(q) + a_{3}(q)M_{12}(q) = 0.
\eeq
The equation (\ref{acond}) is nothing but the condition (\ref{3dim}), which implies that the constraint (\ref{const3}) is integrable by means of an integrating factor. This is a contradiction, since by assumption the constraint (\ref{const3}) is genuinely non-holonomic.

If $\Gamma \not\equiv  0$,  the equations (\ref{eqnsM3}) are inhomogeneous equations. In order that the inhomogeneous equations (\ref{eqnsM3})  may have a solution it is necessary that the quantity $a = (a_{1}, a_{2}. a_{3})$ on its right hand side be orthogonal to any non-trivial  solution (\ref{solM3}) of the homogeneous equation (\ref{homM3}). This also leads to the condition (\ref{acond}) as a necessary condition for the solvability of  Eqns. (\ref{eqnsM3}). This implies the same contradiction as in the homogeneous case.

It has thus been proved that  the d'Alembertian equations of motion (\ref{Lagnh3}) and the variational equations of motion (\ref{CC1}) in the case $N= 3$ can not have coincident solutions if the constraint (\ref{const3}) is genuinely non-holonomic.

\subsection{The cases $N \geq 4$}
\label{dimens4}

I will now consider the linear equations (\ref{eqnsM}) for any finite dimension $N \geq 4$, at some fixed  point $q = (q^{1}, q^{2}, \ldots, q^{N}) \in D_{q}$.

The equations (\ref{eqnsM}), which are necessary consequences of the assumption that d'Alembertian equations of motion (\ref{Lagnh3}) and the variational equations of motion (\ref{CC1}) have coincident solutions, are valid at any given fixed point $q \in D_{q}$.

It is now appropriate to analyse the equations (\ref{eqnsM}) in more detail. Contract Eqns (\ref{eqnsM}) with the quantity $(a^{1}(q), a^{2}(q), \ldots, a^{N}(q))$. This leads to the result
\beq
\label{GamtoR}
\csumAB  a^{A}(q) M_{AB}(q) \dot{q}^{B} = \Gamma ||a(q)||^{2},
\eeq
where the norm $||a(q)||$ is defined by the expression
\beq
\label{norma}
||a(q)||^{2} = \csumA a^{A}(q)a_{A}(q).
\eeq
Let $R(q, \dot{q})$ be defined as follows,
\beq
\label{Def1R}
R(q, \dot{q}) := \csumAB a^{A}(q)M_{AB}(q)\,\dot{q}^{B}.
\eeq
Trading the quantity $\Gamma$ for $R(q, \dot{q})$ in Eqns. (\ref{eqnsM}) one obtains,
\beq
\label{eqnsR}
\sum_{D=1} M^{A}_{~~D}(q)\dot{q}^{D} = \frac{a^{A}(q) R(q, \dot{q})}{||a(q)||^{2}||}, A = 1, 2, \ldots, N,
\eeq
where the free index $A$ is raised for convenience. Contracting Eq. (\ref{eqnsR}) with the quantity 
$\sum_{B=1}^{N}M_{AB}\dot{q}^{B}$, and using the anti-symmetry condition (\ref{antisym}), one gets finally the following expression for the quantity 
$R(q, \dot{q})$ defined in Eq, (\ref{Def1R}),
\beq
\label{Rsquare}
R^{2}(q, \dot{q}) = - ||a(q)||^{2}\, \sum_{B, D=1}^{N} (M^{2})_{BD}(q)\dot{q}^{B}\dot{q}^{D},
\eeq
where
\beq
\label{Msquare}
(M^{2})_{BD}(q) :=  \csumA M_{BA}(q)M^{A}_{~~D}(q), \; B, D = 1, 2, \ldots, N.
\eeq 
From the anti-symmetry condition (\ref{antisym}) follows also that the quantity $M^{2}$ is symmetric in its lower indices,
\beq
\label{Msqsymm}
(M^{2})_{BD}(q)  = (M^{2})_{DB}(q), \; B, D = 1, 2, \ldots, N.
\eeq
Incidentally, Eq. (\ref{Rsquare}) shows that the following quadratic form $Q_{(2)}(x)$ in real variables $x^{A}, A = 1, 2, \ldots, N$, is negative semi-definite,
\beq
\label{defQ2}
Q_{(2)}(x) ;=  \sum_{B, D=1}^{N} \csumA M_{BA}(q)M^{A}_{~~D}(q)x^{B}x^{D} \leq 0.
\eeq
This is made even more explicit by a  reshuffling of the summation indices in Eq. (\ref{defQ2}),
\beq
\label{innerpQ}
Q_{(2)}(x)  = - \sum_{A, C=1}^{N} g^{AC}(q) \left (\sum_{B=1}^{N} M_{AB}(q)x^{B} \right ) \left (\sum_{D=1}^{N} M_{CD})q) x^{D} \right ) \equiv  - ||M(q)x||^{2}.
\eeq
Thus the quadratic form $Q_{(2)}(x)$ vanishes if and only if
\beq
\label{nullQ2}
\csumB M_{AB}(q)x^{B} = 0.
\eeq
The results given above will be used subsequently.

Applying the formula (\ref{innerpQ}) to the expression (\ref{Rsquare}) one gets finally the following useful expression,
\beq
\label{Rfinale}
R(q, \dot{q}) = + ||a(q)|| \, ||M(q)\dot{q}||.
\eeq
There is an ambiguity involving a sign $\pm 1$ in arriving at the expression (\ref{Rfinale}) along the route indicated above. However this sign-ambiguity is of no consequence for the remaining considerations in this paper, so for simplicity I shall stick  to the sign-factor  $+ 1$
in the expression (\ref{Rfinale})  in what follows.

Inserting the expression (\ref{Rfinale}) for $R(q, \dot{q})$ in Eqns. (\ref{eqnsR}) one finds the following equations,
\beq
\label{eqnsRfinal}
\csumB M_{AB}(q)\dot{q}^{B} = \frac{||M(q)\dot{q}||}{||a(q)||}\, a_{A}(q), A = 1, 2, \ldots, N.
\eeq
The tensor equations (\ref{eqnsRfinal}) are  more precise versions of the  equations (\ref{eqnsM}), which  constituted the starting point in the proof of the non-compatibility theorem. 

It is necessary to consider the detailed structure of the solutions $\dot{q}^{A}, A = 1, 2, \ldots, N$, of 
Eqns. (\ref{eqnsRfinal}) in order to proceed with the proof of the non-compatibility theorem enunciated in Subsection \ref{incomp}. For this purpose one needs certain  basic facts concerning skew matrix-like quantities such as the anti-symmetric tensor fields $(M_{AB}(q))$ in the equations above. 

\subsection{Digression on skew matrices $(M_{AB})$}
\label{digrM}

In this subsection I will discuss such properties related to the skew matrix-like quantities 
$M_{AB}(q), \,A, B = 1, 2, \ldots, N$, encountered previously, which are needed for finding  solutions  
 of  Eqns. (\ref{eqnsRfinal}) for $\dot{q}^{A}, A = 1, 2, \ldots, N$. The components  $g_{AB}(q), \; A, B = 1, 2, \ldots, N$ of the metric tensor $g$ enter in an essential way in this discussion.
 
The co-ordinates $q$ enter as parameters in the discussion below, which deals with exclusively  algebraic properties.  In this discussion the co-ordinates $q$ may either considered to correspond  to a fixed point $q_{0} \in D_{q}$, or else  to be confined to an  appropriate reducible sub-domain $D^{0}_{q} \subset D_{q}$, which contains the given fixed point $q_{0}$. Eventually the fixed point $q_{0}$ will be identified with an initial value in co-ordinate space for the d'Alembertian and variational equations of motion under consideration in this paper.
 
Consider the following quadratic form $G_{(2)}(x)$ in certain real variables $x^{A}, A = 1, 2, \ldots, N$,
\beq
\label{G2}
G_{(2)}(x) := \csumAB g_{AB}(q)x^{A}x^{B}.
\eeq
It follows from the assumed properties of the metric $g(q)$, that the  quadratic form $G_{(2)}(x)$ is a real positive definite quadratic form.

I also recall the negative semi-definite quadratic form $Q_{(2)}(x)$ defined above in Eq. (\ref{defQ2}), involving the quantities $(M^{2})_{BD} = (M^{2})_{DB}, \, B, D = 1, 2, \ldots, N$ defined in Eq. (\ref{Msquare}).

It is known that one can effect a simultaneous reduction of two quadratic forms, of which one is a  positive definite form such as the form (\ref{G2}) above,  and the other is merely a real quadratic form, such as the form (\ref{defQ2}) above, to sums of perfect squares \cite{Friedman}. This leads to the following eigen-value problem for eigen-vectors $e^{A}(q), A = 1, 2, \ldots, N$,
\beq
\label{eigenv}
\csumB \left ( (M^{2})_{AB}(q) - \lambda(q) \, g_{AB}(q) \right )e^{B}(q) = 0.
\eeq
The equation (\ref{eigenv}) has non-trivial solutions only if the parameter $\lambda$ satisfies the following eigen-value  equation, which is similar to the characteristic equation for eigen-values in ordinary linear algebra,
\beq
\label{detCC1}
\det  \left ( (M^{2})_{AB}(q) - \lambda(q)\, g_{AB}(q) \right )= 0.
\eeq
Under the conditions described above for the quadratic forms $G_{(2)}(x)$ given in (\ref{G2}), and the quadratic form $Q_{(2)}(x)$ given in (\ref{defQ2}), the equation (\ref{detCC1}) has $N$ eigen-value solutions  $\lambda_{\nu}(q), \nu = 1, 2, \ldots, N$. The corresponding eigen-vectors $e_{\nu}^{~B}(q)$ satisfy the following equations,
\beq
\label{eignu}
\csumB \left ( (M^{2})_{AB}(q) - \lambda_{\nu}(q) \, g_{AB}(q) \right )e_{\nu}^{~B}(q) = 0, \: \;\nu = 1, 2, \ldots, N.
\eeq

It is convenient to consider the following variant of Eqns. (\ref{eignu}), 
\beq
\label{eignulow}
\csumB \left ( (M^{2})_{A}^{~~B}(q) - \lambda_{\nu}(q) \, g_{A}^{~~B}(q) \right )e_{\nu B}(q) = 0, \;\;\nu = 1, 2, \ldots, N.
\eeq
The equations (\ref{eignu}) and (\ref{eignulow}) are perfectly equivalent; the equations (\ref{eignulow}) are obtained by simultaneously raising and lowering the summation indices $B$ inside the equations (\ref{eignu}), which is a straightforward and legitimate operation.

From Eqns. (\ref{eignulow}) follows that the eigen-values $\lambda_{\nu}(q), \nu = 1, 2, \ldots, N$, also satisfy the following variant of Eq. (\ref{detCC1}),
\beq
\label{detCC2}
\det  \left ( (M^{2})_{A}^{~~B}(q) - \lambda_{\nu}(q) \, g_{A}^{~~B}(q) \right )= 0.
\eeq
Here the enumeration of rows and columns in the determinant is as  follows: the lower indices $A = 1, 2, \ldots, N$ enumerate the rows, and the upper indices $B = 1, 2, \ldots, N$ enumerate the columns.

The equation (\ref{detCC2}) has an advantage over Eq. (\ref{detCC1}), namely  that the metric $g$ is only involved in the expressions  $ (M^{2})_{A}^{~~B}(q), A, B = 1, 2, \ldots, N$, in Eq. (\ref{detCC2}). This is due to the fact that  the quantities $g_{A}^{~~B}(q), A, B = 1,2, \ldots, N$  always have the fixed numerical values $0$ or $1$, independently of the co-ordinates $q$,
\beq
\label{gdelta}
 g_{A}^{~~B}(q) :=    \left \{ \begin{array}{ll}
1 & \mbox{if $A = B$}\\
0 & \mbox{if $A \neq B$,}
\end{array}
\right .
\eeq
for $A, B = 1, 2, \ldots, N$. 

It should also be noticed that it follows readily from Eq. (\ref{detCC2}) that the eigen-values  
$\lambda_{\nu}(q), \nu = 1, 2, \ldots, N$, are scalars, {\em i.e.} invariant under co-ordinate transformations.

I finally rewrite Eqns. (\ref{eignulow}) as follows,
\beq
\label{finaleigen}
\csumB  (M^{2})_{A}^{~~B}(q)\, e_{\nu B}(q) = \lambda_{\nu}(q)\, e_{\nu A}(q), \; \nu = 1, 2, \ldots, N, \; A = 1, 2, \ldots, N.
\eeq
It should be noted that the eigen-vectors $e_{\nu A}(q), \; A = 1, 2, \ldots, N$ are orthogonal vectors, as will be shown below.

If the eigen-values $\lambda_{\nu}(q), \nu = 1, 2, \ldots, N$ are non-degenerate, this follows immediately 
from a consideration the equation (\ref{finaleigen}) for two different indices $\mu$ and $\nu$, say. 
Contract  the equations (\ref{finaleigen})  with $e_{\mu}^{~A}(q)$. Interchange $\mu$ and $\nu$ in the equations obtained in this way, and subtract one set of equations from the other set of equations in the pairs of equations obtained by the contraction procedure. Then one obtains,
\beq
\label{ortho}
(\lambda_{\mu}(q) - \lambda_{\nu}(q))(e_{\mu A}(q), e_{\nu B}(q)) = 0,
\eeq
where
\beq
\label{ginnerprod}
(e_{\mu}(q), e_{\nu}(q)) := \csumAB g^{AB}(q)e_{\mu A}(q)e_{\nu B}(q).
\eeq
The abbreviated notation on the left hand side of the expression (\ref{ginnerprod}) for the inner product of vector-quantities such as the eigen-vectors $e_{\nu A}(q), \; A = 1, 2, \ldots, N$, will be frequently used in what follows. 

If the eigen-values are non-degenerate, i.e., if $\lambda_{\mu}(q) \neq \lambda_{\nu}(q)$ when $\mu \neq \nu$, then the orthogonality of the corresponding eigen-vectors $e_{\mu}(q)$ and $e_{\nu}(q)$ follows from the equations (\ref{ortho}). Normalising the eigen-vectors to $1$ in the inner product (\ref{ginnerprod}), one then finally obtains the following ortho-normalisation conditions,
\beq
\label{orthonorm}
(e_{\mu}(q), e_{\nu}(q)) = \delta_{\mu\nu} :=  \left \{ \begin{array}{ll}
1 & \mbox{if $\mu = \nu$}\\
0 & \mbox{if $\mu \neq \nu$,}
\end{array}
\right .
\eeq
for $\mu, \nu = 1, 2, \ldots, N$.

If there are degenerate eigen-values, the ortho-normalisation conditions are imposed as a part of the definition of the corresponding eigen-vectors.  Thus the ortho-normalisation conditions (\ref{orthonorm}) are valid in all cases.

Using the ortho-normalisation conditions (\ref{orthonorm}) one readily obtains the eigen-values
$\lambda_{\nu}(q), \nu = 1, 2, \ldots, N$, from Eqns. (\ref{finaleigen}),
\beq
\label{lambdavalue}
\lambda_{\nu}(q)  =   \csumAB e_{\nu}^{~A}\,(M^{2})_{A}^{~~B}(q)\, e_{\nu B}(q))  \equiv  - ||M(q) e_{\nu}(q)||^{2}, \nu = 1, 2, \ldots, N.
\eeq
In arriving at the final expressions above for the eigen-values  $\lambda_{\nu}, \nu = 1, 2, \ldots, N$, use has been made of the result given in Eq. (\ref{innerpQ}).

From the expressions (\ref{lambdavalue}) follows that all the eigen-values $\lambda_{\mu}, \mu = 1, 2, \ldots, N$ are non-positive,  
\beq
\label{kappa1}
\lambda_{\nu}(q) := - \kappa_{\nu}(q)^{2},
\eeq
where the real quantities  $\kappa_{\nu}(q), \nu = 1, 2, \ldots,, N$, can be chosen to be non-negative without loss of generality. Comparing Eqns. (\ref{lambdavalue}) and (\ref{kappa1}) one then obtains, 
\beq
\label{kappafin}
\kappa_{\nu}(q) =   ||M(q) e_{\nu}(q)||,\;\; \nu = 1, 2, \ldots, N.
\eeq
From Eq. (\ref{kappafin}) one gets an upper bound on the quantities $\kappa_{\nu}, \nu = 1, 2, \ldots, N$, in terms of the norm $||M(q)||$ of the matrix $M(q)$,
\beq
\label{normM}
||M(q)|| = sup_{||x||=1} ||M(q)\,x||.
\eeq
The quantities $\kappa_{\nu}, \nu = 1, 2, \ldots, N$, thus satisfy the following inequalities,
\beq
\label{kappabounds}
0 \leq \kappa_{\nu}(q) \leq ||M(q)||, \; \nu = 1, 2, \ldots, N.
\eeq

From Eq. (\ref{lambdavalue}) follows that an eigen-value $\lambda_{\nu}(q)$ equals zero if and only if
\beq
\label{kerM1}
\csumB M_{AB}(q)\, e_{\nu}^{~B}(q) = 0, \;\; A = 1, 2, \ldots, N.
\eeq
The number of non-vanishing eigen-values is thus determined by the {\em rank} of the matrix 
$M(q)$.  Since $M(q)$ is skew, the rank of $M(q)$ is necessarily an even integer, $2p$, say,
where $p \geq 1$, since $M(q) \not\equiv 0$.

The eigen-values will now be enumerated in such a manner that the last $N-2p$ eigen-values
$\lambda_{\nu}, \nu = 2p+1, \ldots, N$ are equal to zero. Thus,
\beq
\label{kerM}
\csumB M_{AB}(q)\, e_{\nu}^{~B}(q) = 0, \;\;A = 1, 2, \ldots, N;\, \nu = 2p+1, \ldots, N.
\eeq
The subspace $ker(M(q))$ is thus spanned by the eigen-vectors $e_{\nu}(q), \, \nu = 2p+1, \ldots, N$, the components of which have been labelled by  upper or  indices $A = 1, 2, \ldots, N$, in Eqns. (\ref{kerM}).
One may also use lower indices $A, B,\ldots$,  whenever it is convenient, by using the standard rules for lowering or raising such indices with the metric tensor $g_{AB}(q)$ or $g^{AB}(q)$, respectively, where $A, B =  1, 2, \ldots, N$.
Corresponding to the enumeration in Eqns. (\ref{kerM}) one recognises that
\beq
\label{nullkappa}
\kappa_{\nu}(q) = 0, \; \nu = 2p+1, \ldots, N.
\eeq
If $M(q)$ is regular  there are no non-trivial solutions of the equations (\ref{kerM}). This can happen only if $N$ is even and when $2p = N$. 

Generally, for $\nu = 1, \ldots, 2p$, the quantities $\kappa_{\nu}(q)$ are positive,
\beq
\label{poskappa}
\kappa_{\nu}(q) > 0, \;\; \nu = 1, 2, \ldots, 2p.
\eeq

A remark on the classification  of a skew matrix $M(q)$ according to its rank may be in order. Since 
$M(q)$ depends on $q$, also the rank $2p$ is in principle dependent on $q$. If considerations related to a skew matrix  $M(q)$ involving  only one prescribed fixed point $q_{0}$, say,  no problems can occur by the omission of the argument $q$ in the quantity $p$. This notation is used here for the sake of simplicity.  Strictly speaking, however, one should use the notation $p(q)$ in stead of only $p$. The rank $2p(q)$ takes values in the set of even positive integers, and therefore the rank $2p(q)$  ought not to change from its value at a prescribed point $q_{0}$ if one considers neighbouring points $q$ sufficiently near the prescribed point $q_{0}$. This also justifies the omission of the argument $q$ from the symbol $2p$, which then can be considered to be a valid characterisation of $M(q)$ at least  in an appropriate sufficiently small domain containing the prescribed  point $q_{0}$ as an interior point.

I will now construct a new basis in the $N$-dimensional linear space spanned by the eigen-vectors
$e_{\nu}(q), \,\nu = 1, 2, \ldots, N$. This basis is used to solve the important  equations (\ref{eqnsRfinal}).
The construction parallels closely a similar construction made by Greub \cite{Greub} for skew mappings in ordinary Euclidean space.

Define vectors $b_{2\nu-1}(q), \, \nu = 1, \ldots, p$, as follows,
\beq
\label{Defbodd}
b_{2\nu-1}^{~~~~~A}(q) := e_{\nu}^{~A}(q), \, \nu = 1, \ldots, p\; ; A = 1, 2, \ldots, N.
\eeq
Similarly, let  vectors $b_{2\nu}(q), \, \nu = 1, \ldots, p$, be defined by the equations,
\beq
\label{Defeven}
b_{2\nu A}(q) := \kappa_{\nu}(q)^{-1}\csumB\, M_{AB}(q)b_{2\nu-1}^{~~~~~B}(q), \, \nu = 1, \ldots, p\; ; A = 1, 2, \ldots, N.
\eeq
In view of the conditions (\ref{poskappa}) the definition is a possible definition.

I  rewrite the eigen-value equations (\ref{finaleigen}) with $- \kappa_{\nu}(q)^{2}$ in stead of $\lambda_{\nu}(q)$, thus
\beq
\label{finalkap}
\csumB  (M^{2})_{A}^{~~B}(q)\, e_{\nu B}(q) = - \kappa_{\nu}(q)^{2}\,(q)\, e_{\nu A}(q), \; \nu = 1, 2, \ldots, N, \; A = 1, 2, \ldots, N.
\eeq

It is now a simple matter to show that
\beq
\label{beigen}
\csumB  (M^{2})_{A}^{~~B}(q)\, b_{\mu B}(q) =   - \kappa_{\nu}(q)^{2} \, b_{\mu A}(q), \; \mu = 1, 2, \ldots, 2p, \; A = 1, 2, \ldots, N.
\eeq

When $2p < N$ one also defines, 
\beq
\label{kerM3}
b_{\nu A}(q) := e_{\nu A}(q), \; \nu = 2p+1, \ldots, N,\; , A = 1, 2, \ldots, N.
\eeq
Then,
\beq
\label{nulleigen}
\csumB  (M^{2})_{A}^{~~B}(q)\, b_{\nu B}(q) = 0, \; \nu = 2p+1, \ldots, N \;  ;A = 1, 2, \ldots, N.
\eeq

From the definition (\ref{Defeven}) follows that
\beq
\label{Meqodd}
\csumB\, M_{AB}(q)b_{2\nu-1}^{~~~~~B}(q) = \kappa_{\nu}(q) b_{2\nu A}(q), \;, \nu = 1, \ldots, p\; : \; A = 1, 2, \ldots, N.
\eeq
Using the equations (\ref{beigen}) it is also simple to show that
\beq
\label{Meqeven}
\csumB\, M_{AB}(q) b_{2\nu}^{~~~B}(q) = -  \kappa_{\nu}(q) b_{2\nu-1 A}(q), \;, \nu = 1, \ldots, p\; : \; A = 1, 2, \ldots, N.
\eeq
It is finally stated without proof, that the basis vectors $b_{\mu}(q), \mu = 1, 2, \ldots, N$, defined above, are orthonormal in the
 inner product (\ref{ginnerprod}),
\beq
\label{bortho}
(b_{\mu}(q), b_{\nu}(q)) = \delta_{\mu \nu}, \; \mu, \nu = 1, 2, \ldots, N.
\eeq
The proof of the ortho-normalisation conditions (\ref{bortho}) is fairly simple, and is left to the interested reader.

The vectors $b_{\mu},\, \mu = 1, 2, \ldots, N$ thus also form an orthonormal basis in the space spanned by the eigen-vectors $e_{\mu}, \, \mu = 1, 2, \ldots, N$ of the matrix $M^{2}(q)$.

Incidentally, the results given above imply that there are only $p$ non-zero eigen-values $\lambda_{\nu}(q) = -\kappa_{\nu}(q)^{2}$ related to the matrix $M^{2}(q)$ in stead of $2p$ such values. This means that the non-zero eigen-values of $M^{2}(q)$ are at least two-fold degenerate.

\subsection{The solutions  $\dot{q}$}
\label{solutionsdotqC}
I now return to the important equations (\ref{eqnsRfinal}). These equations are valid for $q \in D_{q}$, but will later only be considered at a fixed point  $q_{0}$, which is  identified as the initial value point for the d'Alembertian and variational equations of motion (\ref{Lagnh3}) and (\ref{CC1}), respectively. The initial velocity at $q_{0}$ will  be denoted by $\dot{q}_{0}$.

For simplicity of notation, I will for the time being consider the equations at an arbitrary  fixed point $q \in D_{q}$.

Contracting the equations (\ref{eqnsRfinal}) with an arbitrary basis-vector $b_{\nu}(q)$, with $\nu$ in the range $2p+1, \ldots, N$, one obtains,
\beq
\label{bRcond}
(b_{\nu}(q), a(q))\,||M(q)\dot{q}||  = 0, \; \nu = 2p+1, \ldots, N.
\eeq
From Eq. (\ref{bRcond}) follows that {\em either},
\beq
\label{Rzero}
||M(q)\dot{q}||  = 0,
\eeq
{\em or},
\beq
\label{Rnotzero}
||M(q)\dot{q}||  \neq 0,
\eeq
in which case the following conditions must also be fulfilled,  
\beq
\label{altRcond}
(b_{\nu}(q), a(q)) = 0, \; \nu = 2p+1, \ldots, N.
\eeq

\subsubsection{The case $||M(q)\,\dot{q}|| = 0$.}

Consider first the case (\ref{Rzero}) with $2p < N$. Thus $\dot{q} \in ker(M(q)$, or,  equivalently,
\beq
\label{solqdothom1}
\dot{q}^{A} = \sum_{\nu=2p+1}^{N} \gamma_{\nu}\, b_{\nu}^{~A}(q), \;A = 1, 2, \ldots, N.
\eeq
The parameters $\gamma_{\nu}, \nu = 2p+1, \ldots, N$, in the expressions (\ref{solqdothom}) are restricted by the non-holonomic condition (\ref{1nonhol}). Thus,
\beq
\label{condam1}
0 = (a(q), \dot{q}) = \sum_{\nu=2p+1}^{N} \gamma_{\nu}\, (a(q), b_{\nu}(q)).
\eeq

I now consider the results above at the fixed initial value point $q_{0}$.

The equation (\ref{solqdothom1}) then reads as follows.
\beq
\label{solqdothom}
\dot{q_{0}}^{A} = \sum_{\nu=2p+1}^{N} \gamma_{0 \nu} \,b_{\nu}^{~A}(q_{0}), \;A = 1, 2, \ldots, N.
\eeq
The parameters $\gamma_{0 \nu}, \, \nu = 2p+1, \ldots,, N$, are restricted by the condition (\ref{condam1}) evaluated at $q_{0}$,
\beq
\label{condam}
0 = (a(q_{0}), \dot{q}_{0}) = \sum_{\nu=2p+1}^{N} \gamma_{0 \nu}\, (a(q_{0}), b_{\nu}(q_{0})).
\eeq
However, the condition (\ref{condam}) is nugatory if the vector $a(q_{0})$ is orthogonal to $\ker(M(q))$.
There are  thus at most $N-2p$ parameters $\gamma_{0 \nu}$ available for the $N-1$ independent initial values obtainable from the initial velocity components $\dot{q}_{0}^{A}, A = 1, 2, \ldots, N$. This means that the system of equations (\ref{solqdothom}), considered as equations for the determination of the parameters $\gamma_{0 \nu}, \nu = 2p+1, \ldots, N$ is  over-determined. These equations can be consistent  only if the given initial values $\dot{q}_{0}^{A}, A=1, 2,\ldots, N$ satisfy at least $(N-1) - (N-2p) = 2p-1 \geq 1$ consistency conditions, in addition to the constraint (\ref{initrestr}).  This contradicts the requirement that  one should be able to impose general initial values $\dot{q}_{0}^{A}, \, A = 1, 2, \ldots, N$, for the velocities $\dot{q}^{A}, A = 1, 2, \ldots, N$,  except for the restriction (\ref{initrestr}), which follows from the non-holonomic condition (\ref{1nonhol}) at the initial value point $q_{0}$.

It has thus been shown that  that the initial values $\dot{q}_{0}^{A}, A = 1, 2, \ldots, N$, which satisfy the condition  implied by the non-holonomic constraint (\ref{1nonhol}) evaluated at $q_{0}$, can not be  freely chosen, apart from the restriction (\ref{initrestr}), if one demands that the d'Alembertian equations (\ref{Lagnh3}) and the variational equations (\ref{CC1}) should have coincident solutions. 

There is still the possibility that $ker(M(q))$ contains only the zero vector. This is the case if the matrix $M$ is regular, which is possible only if $N$ is an even integer. Then there is only the trivial solution $\dot{q}_{0} = 0$, which naturally is  in conflict with the requirement that it should be possible to choose the initial values $\dot{q}_{0 A}, A = 1, 2, \ldots, N$,  freely, except for the condition implied by the non-holonomic constraint (\ref{1nonhol}) at the point $q_{0}$.

\subsubsection{The case $||M(q)\, \dot{q}|| \neq 0$.}
\label{importanteM}

It has been shown above that the  case (\ref{Rzero}) necessarily leads to contradictions. It remains to consider the   case (\ref{Rnotzero}). Then the conditions (\ref{altRcond}) must also be in force. 

In view of the conditions (\ref{altRcond}),  the vector $a(q)$ must lie in the subspace spanned by the basis-vectors $b_{\mu}(q)$, for $\mu = 1, 2, \ldots, 2p$,
\beq
\label{repra}
a(q) = \sum_{\mu=1}^{p}\left \{ \omega_{2\mu-1}\,b_{2\mu-1}(q) + \omega_{2\mu}\,b_{2\mu}(q)\right \},
\eeq
where for later convenience we have separated the sum in the representation (\ref{repra}) into two parts, as sums over even and odd indices, respectively. The coefficients $\omega_{2\mu-1}$ and $\omega_{2\mu}$, respectively, for $\mu = 1, \ldots, p$, in the representation (\ref{repra}), are determined by the vector $a(q)$.

I will  now consider such solutions $z$, say, of the equations (\ref{eqnsRfinal}) which are in the linear span of the basis-vectors $b_{\mu}(q)$ for $\mu = 1, 2, \ldots, 2p$. Thus,
\beq
\label{reprz}
 z = \sum_{\mu=1}^{p} \left \{ \alpha_{2\mu-1}b_{2\mu-1}(q) + \alpha_{2\mu}b_{2\mu}(q) \right \}.
\eeq
Inserting  $\dot{q} = z$ in Eqns. (\ref{eqnsRfinal}) one obtains the following equations for the components $z^{A}, A = 1, 2, \ldots, N$,
\beq
\label{eqnsMz}
\sum_{B=1}^{N} M_{AB}(q)z^{B} =  \frac{||M(q)\,z||}{||a(q)||}\,a_{A}(q), \;A = 1, 2, \ldots, N.
\eeq
Contracting the equations  (\ref{eqnsMz}) with the basis-vector components $b_{2\nu}^{~~A}(q)$
and using the equations (\ref{Meqeven}),  one obtains the following result,
\beq
\label{oddcompz}
\kappa_{\nu}(q) (b_{2\nu-1}(q), z) = \frac{||M(q)\,z||}{||a(q)||}\, (b_{2\nu}(q), a(q)), \;\nu = 1, \ldots, p.
\eeq
Likewise, contracting the (\ref{eqnsMz}) with the basis vector components $b_{2\nu-1}^{~~~~~A}(q)$, and using the equations (\ref{Meqodd}), one finds a similar expression,
\beq
\label{evencompz}
\kappa_{\nu}(q) (b_{2\nu}(q), z) = -  \frac{||M(q)\,z||}{||a(q)||}\, (b_{2\nu-1}(q), a(q)), \; \nu = 1, \ldots, p.
\eeq
The equations (\ref{oddcompz}) and (\ref{evencompz}) mean that  the equations (\ref{eqnsRfinal}) for the vector $z$ have been solved in the sense  that the equations (\ref{oddcompz}) and (\ref{evencompz}) express all  the components  $\alpha_{2\nu-1}$ and $\alpha_{2\nu}$, respectively, in the representation (\ref{reprz}) for $z$,  in terms of one unknown scalar quantity, namely $||M(q)\,z||$, and the known components $\omega_{2\nu-1}$ and $\omega_{2\nu}$, respectively, in the representation (\ref{repra}) for the vector $a(q)$.

The result obtained above can be summarised as follows,
\beq
\label{zfinale1}
z^{A}=   \frac{||M(q)\, z||}{||a(q)||} \sum_{\nu=1}^{p} \kappa_{\nu}^{-1}(q) \left \{(b_{2\nu}(q), a(q))\,b_{2\nu-1}^{~~~~~ A}(q) - (b_{2\nu-1}(q), a(q))\,b_{2\nu}^{~~ A}(q) \right \}.
\eeq
It should  noted that the vector $z$ defined by Eqns. (\ref{zfinale1}) automatically fulfils the following orthogonality condition,
\beq
\label{z0ortoa1}
(z, a(q)) = 0.
\eeq
So far only the solutions $z^{A}, A = 1, 2, \ldots, N$,  of the equations (\ref{eqnsRfinal}) in the linear span of the basis vectors  $b_{\mu}(q), \mu = 1, \ldots, 2p$, have been considered. The general solution $\dot{q}$ of Eqns. (\ref{eqnsRfinal}) is  a sum of the solution $z$ given above in Eqns. (\ref{zfinale1}), and a general solution $h$, say,  of the corresponding homogeneous equations. Thus,
\beq
\label{H0}
h^{A} := \sum_ {\nu=2p+1}^{N} \gamma_{\nu}\, b_{\nu}^{~A}(q),
\eeq
where the parameters $\gamma_{\nu}, \nu = 2p+1, \ldots, N$ are free parameters. 

The general solutions of Eqns. (\ref{eqnsRfinal}) for the components $\dot{q}^{A}$ are  the  following,
\beq
\label{qsol1}
\dot{q}^{A} := z^{A} + h^{A}, \; A = 1, 2, \ldots, N
\eeq
It should be noted that
\beq
\label{Cmnorm}
||M(q)\dot{q}|| = ||M(q)(z+ h)|| = ||M(q)\,z||.
\eeq
One thus finally obtains the following expressions for the general solutions of Eqns. (\ref{eqnsRfinal}),
\bea
\label{qdotfinale1}
\dot{q}^{A} & =  & \frac{||M(q)\,\dot{q}||}{||a(q)||} \sum_{\nu=1}^{p} \kappa_{\nu}^{-1}(q) \left \{(b_{2\nu}(q), a(q))\,b_{2\nu-1}^{~~~~~ A}(q) - (b_{2\nu-1}(q), a(q))\,b_{2\nu}^{~~ A}(q) \right \} \nonumber \\
& + &  \sum_ {\nu=2p+1}^{N} \gamma_{\nu}\, b_{\nu}^{~A}(q), \;A = 1, 2, \ldots, N.
\eea
In view of the conditions (\ref{altRcond})  and (\ref{z0ortoa1}), one finds that the final expressions
(\ref{qdotfinale1}) for the components  $\dot{q}_{0}^{A}, \, A = 1, 2, \ldots, N$, automatically satisfy the non-holonomic condition (\ref{1nonhol}).

I now apply the results above for the initial value $\dot{q}_{0}$ at  $q_{0}$. 

It follows from  Eqns. (\ref{qdotfinale1}) that
\bea
\label{qdotfinale}
\dot{q}_{0}^{A} & =  & \frac{||M(q_{0})\,\dot{q}_{0}||}{||a(q_{0})||} \sum_{\nu=1}^{p} \kappa_{\nu}^{-1}(q_{0}) \left \{(b_{2\nu}(q_{0}), a(q_{0}))\,b_{2\nu-1}^{~~~~~ A}(q_{0}) - (b_{2\nu-1}(q_{0}), a(q_{0}))\,b_{2\nu}^{~~ A}(q_{0}) \right \} \nonumber \\
& + &  \sum_ {\nu=2p+1}^{N} \gamma_{0 \nu}\, b_{\nu}^{~A}(q_{0}), \;A = 1, 2, \ldots, N.
\eea
When the initial velocity components $\dot{q}_{0}^{A}, A = 1, 2, \ldots, N$ are given,  the only unknown quantities in the equations (\ref{qdotfinale}) are the $N-2p$ parameters $\gamma_{0 \nu}, \nu = 2p+1, \ldots, N$.

When $2 \leq 2p < N$, there are altogether  $N-1$ independent equations among the $N$ equations in (\ref{qdotfinale}),  to be solved for the $N-2p$ parameters $\gamma_{0 \nu}, \nu = 2p+1, \ldots, N$, when the components $\dot{q}_{0}^{A}, A = 1, 2, \ldots, N$ are given. This is an over-determined system of equations for the parameters  $\gamma_{0 \nu}, \nu = 2p+1, \ldots, N$, which implies that there must be at least $(N-1) - (N-2p) = 2p-1 \geq 1$ consistency conditions among the components $\dot{q}_{0}^{A}, A = 1, 2, \ldots, N$, in addition to the constraint (\ref{initrestr}). This contradicts  the requirement that it should be possible to choose the initial velocity $\dot{q}_{0}$ freely, except for the constraint (\ref{initrestr}).

For even $N$ one can have $2p = N$. Then the solutions for $\dot{q}_{0}^{A}, \, A = 1,2,\ldots, N$ 
are as follows,
\beq
\label{qdotfinale2}
\dot{q}_{0}^{A}  =   \frac{||M(q_{0})\,\dot{q}_{0}||}{||a(q_{0})||} \sum_{\nu=1}^{p} \kappa_{\nu}^{-1}(q_{0}) \left \{(b_{2\nu}(q_{0}), a(q_{0}))\,b_{2\nu-1}^{~~~~~ A}(q_{0}) - (b_{2\nu-1}(q_{0}), a(q_{0}))\,b_{2\nu}^{~~ A}(q_{0}) \right \} 
\eeq
The $N$ components  $\dot{q}_{0}^{A}, \; A = 1, 2, \ldots, N$ in the equations (\ref{qdotfinale2}),  are  given in terms of one undetermined scalar quantity $||M(q_{0})\,\dot{q}_{0}||$, in addition to the other known quantities in (\ref{qdotfinale2}). This means that there are $N-1$ constraints among the components   
$\dot{q}_{0}^{A}, \; A = 1, 2, \ldots, N$, which clearly contradicts  the requirement that it should be possible to choose the initial velocity $\dot{q}$ freely, except for the constraint (\ref{initrestr}).

 In the case $||M(q_{0})\,\dot{q}_{0}|| \neq 0$, it has thus been shown that there are  at least $2p-1$ consistency conditions on the velocity components $q_{0}^{A}$. There are also $N-2p$ conditions of the form (\ref{altRcond}) evaluated at $q_{0}$, namely the following conditions,
\beq
\label{altRcond2}
(b_{\nu}(q_{0}), a(q_{0})) = 0, \; \nu = 2p+1, \ldots, N, 
\eeq
Taken together there are thus altogether at least $(2p-1) + (N-2p) = N-1$ conditions on the initial values $(q_{0}^{A}, \dot{q}_{0}^{A}), A = 1, 2, \ldots, N$. This is a set of stronger  conditions than the consistency conditions involving only the velocity components $\dot{q}_{0}^{A}, A = 1, 2, \ldots, N$.

It has thus been shown also in the case  $||M(q)_{0})\dot{q}_{0}|| \neq 0$, that  that the initial values $\dot{q}_{0}^{A}, A = 1, 2, \ldots, N$, which satisfy the condition (\ref{initrestr})  implied by the non-holonomic constraint (\ref{1nonhol}) evaluated at $q_{0}$, have to satisfy additional consistency conditions, if one demands that the d'Alembertian equations (\ref{Lagnh3}) and the variational equations (\ref{CC1}) should have coincident solutions. 

\subsection{Conclusion of the proof of the incompatibility theorem}

The starting point in of the beginning of the proof in Subsection \ref{preliminary},  is the following set of equations,
\beq
\label{eqnsM2}
\sum_{B=1}^{N} M_{AB}(q)\,\dot{q}^{B} = \Gamma\,a_{A}(q), \; A = 1, \ldots, N,
\eeq
where $\Gamma$ is a scalar  parameter. The equations (\ref{eqnsM2}) are necessary consequences of the assumption that the variational equations (\ref{CC1}) and the d'Alembertian equations (\ref{Lagnh3}) have  coincident solutions.

It was shown in Subsection \ref{dimension3} in the case $N=3$, that the equations (\ref{eqnsM2}) do not have solutions $\dot{q}^{A}, \, A = 1,2,3$, satisfying the required constraint  (\ref{1nonhol}), if this constraint is genuinely non-holonomic, {\em i.e.},  neither integrable as such, nor integrable by means of an integrating factor. 

Thus the  variational equations (\ref{CC1}) and the d'Alembertian equations (\ref{Lagnh3}) are not compatible when $N=3$.

When $N\geq4$, the equations (\ref{eqnsM2}) were transformed into the following form in Subsection
\ref{dimens4},
\beq
\label{eqnsRfinalC2}
\sum_{B=1}^{N} M_{AB}(q)\dot{q}^{B} = \frac{||M(q)\dot{q}||}{||a(q)||}\, a_{A}(q), A = 1, 2, \ldots, N,
\eeq
where the norms $||\ldots||$ occurring above are defined by an appropriate  metric tensor $g$.

For $N\geq4$  the variational equations of motion (\ref{CC1}),  and the d'Alembertian equations of motion (\ref{Lagnh3}), respectively, have been considered to be initial value problems.

The implications of Eqns. (\ref{eqnsRfinalC2}) for the initial values $q_{0}$ and $\dot{q}_{0}$  were analysed in Subsection \ref{solutionsdotqC}. It was shown that the equations (\ref{eqnsRfinalC2}) imply that the initial values $\dot{q}_{0}^{A}, \, A = 1,2,\ldots, N$, for the solutions of the variational equations (\ref{CC1}) and d'Alembertian equations (\ref{Lagnh3}), which satisfy the required  constraint (\ref{1nonhol}), will have to satisfy additional consistency conditions if the constraint (\ref{1nonhol}) is genuinely non-holonomic.

Thus in the cases $N\geq4$,  the  d'Alembertian and variational equations of motion are incompatible in the sense that in the case of a genuinely non-holonomic constraint, they can not have coinciding solutions with  initial velocities $\dot{q}_{0}^{A}, \, A = 1, 2, \ldots, N$, which satisfy the required constraint (\ref{1nonhol}) at $q_{0}$, unless these initial velocities satisfy additional consistency conditions. 
 
This concludes the proof of the incompatibility theorem stated in Subsection \ref{incomp}.

\section{Discussion}

In this paper I have refined and completed an earlier proof \cite{CCTRJMP} of the incompatibility of the variational equations of motion (\ref{CC1}) and the d'Alembertian equations of motion (\ref{Lagnh3}).
It was observed in Ref. \cite{CCMN70}, that the proof in Ref. \cite{CCTRJMP} is not valid in the cases 
$p = 1$ and $N\geq4$, since the conditions of consistency derived in the proof in Ref. \cite{CCTRJMP} are nugatory when $p=1$ and $N\geq4$. This  circumstance  has now been taken into account in the present proof, in which more precise consistency conditions have been derived than in the earlier proof. 

In the present proof the formulation respects explicitly the desirable circumstance that the geometry of configuration space need not be Euclidean, but rather {\em e.g.} Riemannian. The tensor machinery 
necessary in this case, which was  developed in Subsection \ref{dimens4}, and in particular in Subsection \ref{digrM}, may seem a little heavy. It is a price one has to pay for the desire to handle a Riemannian configuration space using a classical tensor formulation.

The proof in Ref. \cite{CCTRJMP} avoided using a full-fledged tensor formalism, by using essentially only local analysis related to the fixed initial point $q_{0}$ in configuration space for the cases 
$N\geq4$, and by implicitly using so-called normal- or geodesic co-ordinates \cite {Eisenh} at that point.  This  means that one was permitted to use a Cartesian metric at the fixed point $q_{0}$ in Ref. \cite{CCTRJMP}.

The incompatibility theorem  proved in this paper does not mean that one is always unable to associate some variational principle with systems with non-holonomic constraints.  It has only been proved that
the particular variational equations (\ref{CC1}) are incompatible with the d'Alembertian equations of motion when $N\geq4$, in the sense that these equations do not have solutions with general initial velocities, if the  homogeneous constraint used in this paper is genuinely non-holonomic. For  $N=3$ a stronger result was proved, namely that the equations of motion in question do not  have  coinciding solutions at all, if the  constraint is genuinely non-holonomic.

Conversely, it is by no means guaranteed that the variational equations (\ref{CC1}), and the d'Alembertian equations (\ref{Lagnh3})  have coinciding solutions, if the initial values 
$\dot{q}_{0}^{A}, \, A = 1, 2, \ldots, N$, satisfy those partly explicit and partly implicit consistency conditions which have been derived in this paper. For instance, the consistency consistency conditions involving only the initial velocities do by no means exhaust all the implications of the equations (\ref{eqnsRfinalC2}) for the initial values $q_{0}$ and $\dot{q}_{0}$, not to mention additional consistency requirements  which come into play when one takes into account the full information residing in the equations of motion. An example of reasonably simple yet intricate additional consistency conditions on the initial values $q_{0}$ was given for an important class of solutions of the equations 
(\ref{eqnsRfinalC2}) in Subsection \ref{importanteM}.

\vspace*{2cm}  

\noindent
\large
{\bf Acknowledgements}
\normalsize

I am indebted to the organisers of the  "4th International Young Researchers Workshop on Geometry, Mechanics and Control", which took place at the University of Ghent in January 11-13, 2010, and in particular to Dr. Tom Mestdag, for inviting me to this inspiring meeting.  I would also like to thank my friend Professor Jerry Segercrantz for an illuminating discussion when I was finalising the manuscript of this paper.

\end{document}